\documentstyle[12pt]{article}
\begin{document}

\title{Decoherence from internal degrees of freedom for clusters of mesoparticles:
a hierarchy of master equations }
\author{J. C. Flores}
\date{Universidad de Tarapac\'a \\
Departamento de F\'{\i}sica\\
Casilla 7-D, Arica \\
Chile}
\maketitle

\baselineskip=17pt

A mesoscopic evolution equation for an ensemble of mesoparticles follows
after the elimination of internal degrees of freedom. If the system is
composed of a hierarchy of scales, the reduction procedure could be worked
repeatedly and the characterization of this iterating method is carried-out.
Namely, a prescription describing a discrete hierarchy of master equations
for the density operator is obtained. Decoherence follows from the
irreversible coupling of the system, defined by mesoscopic variables, to
internal degree of freedom. We discuss briefly the existence of systems with
the same dynamics laws at different scales. We made an explicit calculation
for an ensemble of particles with internal harmonic interaction in an
external anharmonic field. New conditions related to the semiclassical limit
for mesoscopic systems (Wigner-function) are conjectured.

$$
$$


\newpage

\begin{section}*{ I.- Introduction : reduction technique } 
\end{section}

The study of models allowing a unified description of microscopic and
macroscopic physical systems has a long history. The problem is related to
microscopic superposition of states and its non-occurrence at macroscopic
scales. Interesting responses and different proposition can be found in
references [1-18] which are related to coherence destruction by different
approaches to the macroscopic level. Many of these theories are related to
the original ideas developed by Landau [1] respect to the high density of
states for macroscopic object, and high sensibility to external
perturbation. Thus, any small perturbation produces an undefined
(macroscopic) state and then mixture (i.e. decoherence). Nevertheless, it is 
$naive$ to think that the physics between elementary particles and
macroscopic objects, for instance like to macro-molecules, can be described
only for one mesoscopic theory. This is the case of the DNA-macromolecule
which, in a first level, is composed of interacting atoms and finally
becomes responsible for transmission of genetic information in a biological
level. Other examples are some biological composites like to the
hierarchical organization of tendon, bone, mollusk shell, synthetic
composites [19,20] and others.

$$
{} 
$$

The scope of this paper is the study of an iterating systematic procedure at
different discrete scales of perception. The idea is simple, we start with
an ensemble of elementary particles forming clusters around its
mass-centers, then we eliminate the internal degrees of freedom. In this way
the dynamics law for this reduced system is obtained and the procedure could
be worked again to the next level. 
$$
{} 
$$

The general assumptions to construct the hierarchy of master equations are:

\begin{quotation}
$$
{} 
$$
(i) The system is composed of a hierarchy of scales, or levels, where we can
recognize different architectures of clusters. 
$$
{} 
$$
(ii) There are intercorrelation between levels. Specifically, a given set of
clusters make-up a cluster in the next scale. 
$$
{} 
$$
(iii) The dynamical requirement (equations of motion), in a given scale, is
depending onto the above one.

$$
{} 
$$
\end{quotation}

Assumption (i) and (ii) are related to geometrical aspects, and (iii)
possibilities the obtention of dynamical laws from more fundamentals scales.
In fact, (iii) is related to the usual belief that phenomenological laws can
be explained from fundamental models.

$$
{} 
$$

The paper is organized as follows. In this first section, we discuss briefly
the reduction method applied to an ensemble of generic systems. In section
II we deal with the iterating method and finding the master equation
describing the dynamics, at different discrete scales, for interacting
mesoparticles. In III some examples are briefly studied and we investigate
the question related to the existence of invariant-systems under reduction
procedure. In section IV, we use the Wigner function to explore the
semiclassical limit for the quantum evolution equation. The mesoscopic term,
related to the internal degree of freedom, requires new conditions aside the
usual one related to the optical geometric limit. Conclusions and
discussions are presented in the ending section. 
$$
{} 
$$

Now we reviewed briefly the reduction procedure [21-27] which will be used
in the next section. Consider the interaction between a system $S$ and other 
$R$, with many degrees of freedom, and the evolution equation for the
complete system

\begin{equation}
\partial _t\rho ={\cal L}\rho .
\end{equation}
We assume that the Liouville-von Neumann operator is decomposed like 
\begin{equation}
{\cal L}={\cal L}_S+{\cal L}_R+{\cal L}_I
\end{equation}
where ${\cal L}_I$ denotes the interaction term. Consider the projector
operator $P$ acting onto the total density operator $\rho $ (or the space
density distribution in the classical case) : $P\rho =\rho _R^eTr_{_R}(\rho )
$ where $\rho _R^e$ denotes the equilibrium state of the system $R$. In the
classical case, the partial-trace-operation is replaced by an integral over
the phase-space of $R$. As usually, the projection operator $P$ satisfies
[22] : 
\begin{equation}
P{\cal L}_R={\cal L}_RP=0,\quad P{\cal L}_S={\cal L}_SP,\quad P{\cal L}_IP=0.
\end{equation}
In this way from the evolution equation (1), for the complete system, and
projecting on the space spanned by $P$ and $Q=1-P$, one obtains the equation
for $S$ : 
\begin{equation}
\partial _t\rho _S={\cal L}_S\rho _S+Tr_{_R}{\cal L}_I\int_{0^{-}}^td\tau e^{%
{\cal L}\tau }{\cal L}_I\rho _R^e\rho _S(t-\tau ).
\end{equation}
To obtain the above equation the usual initial condition $\rho (0)=\rho
_R^e\rho _S$ was assumed. Equation (4) is exact and cumbersome because the
integral term is dependent on the history of $S$. Expanding to second order
in the interaction term one obtains the equation 
\begin{equation}
\partial _t\rho _S={\cal L}_S\rho _S+\int_{0^{-}}^td\tau \langle {\cal L}_I%
{\cal L}_I(\tau )\rangle _{_R}\rho _S(t),
\end{equation}
where $\langle \circ \rangle _{_R}$ denotes the partial trace operation $%
Tr_{_R}(\circ \rho _R^e),$ and the symbol `$\circ $' means an element of the
space of operators. Assuming the usual memory loss property or Markov
approximation : 
\begin{equation}
\langle {\cal L}_I{\cal L}_I(\tau )\rangle _{_R}=\gamma \delta (\tau ){\cal L%
}_{IS}^2,
\end{equation}
with $\gamma $ a positive parameter and ${\cal L}_{IS}$ an operator acting
on $\rho _S$, then we obtain from (5) and (6), the evolution equation for $S$
\begin{equation}
\partial _t\rho _S={\cal L}_S\rho _S+\gamma {\cal L}_{IS}^2\rho _S.
\end{equation}
The explicit verification of the properties : $\rho _S=\rho _S^{\dagger }$, $%
Tr\rho _S=1$ and $\rho _S>0$ (positivity) must be carried-out always. Remark
that currently an equation like (7) is related to decoherence. Specifically,
the reservoir $R$ changes any system pure states to mixed. An approximated
case, where ${\cal L}_{IS}={\cal L}_S$, with decoherence and without
dissipation can be found in [8,9]. For a criticism to the reduction
procedure see for instance [10,12] where a completely integrable system was
considered. Dissipating effects are treated in [4] (non-linear equation)
where other techniques were considered.

$$
{} 
$$

\begin{section}*{ II.- Reduction procedure: hierarchiy of master equations}
\end{section}
$$
{} 
$$

In this section we use the reduction procedure sketched in section I,
including a coordinate change to the center of mass, and we obtain the
evolution operator at the next scale. Namely, we formalize the procedure $%
{\cal L}^{(n)}\to {\cal L}^{(n+1)}$. So at scale $n$, we start with $N^{(n)}$
interacting particles in an external field. This set contains an
architecture of $N^{(n+1)}$ clusters which are recognized by using physical
constraints (assumption (i)). For instance they could be $N^{(n+1)}$
molecules in an electric field or interacting macromolecules. After the
cluster recognition, we consider a coordinate change to the center of mass
(assumption (ii)), eliminating the internal degrees of freedom for every
cluster (tracing-out-technique). 
$$
{} 
$$

Formally at scale $n$, the equation of motion for the density distribution
in the classical case, or the density operator in the quantum case, is 
\begin{equation}
\partial _t\rho ^{(n)}={\cal L}^{(n)}\rho ^{(n)}, 
\end{equation}
where ${\cal L}^{(n)}$ denotes a linear operator constructed, for instance,
by elementary Liouvillian (or von Neumann) operators like ${\cal L}_f\circ
=\{f,\circ \},$ (or ${\cal L}_f\circ =(1/i\hbar )[f,\circ ]$). Where the
symbol$\{\circ ,\circ \}$ ( or $[\circ ,\circ ]$) stands for the usual
Poisson brackets (commutator). 
$$
{} 
$$

In (8), the index $n$ becomes related to the discrete scale, and the idea is
to obtain the evolution equation, at next scale, by constructing the new
operator ${\cal L}^{(n+1)}$ from the dynamics at scale $n$ (assumption
(iii)). In this way, the reduction procedure has technically two steps:

$$
{} 
$$

(a) A coordinate change to the center-of-mass of every cluster.

(b) Elimination of internal degrees of freedom by assuming

loss-memory-effects (i.e. internal complexity).

$$
{} 
$$

Therefore, the reduction $N^{(n)}\to N^{(n+1)}$ gives an equation similar to
(8), where the Liouville-von Neumann operator is determined using steps (a)
and (b). To find the new evolution operator, we assume the decomposition : 
\begin{equation}
{\cal L}^{(n)}={\cal L}_K^{(n)}+{\cal L}_V^{(n)}. 
\end{equation}
Namely, a kinetic part depending on momentum and other depending on position.

$$
{} 
$$

To begin, we explicitly consider  the first reduction procedure because it
contains all the basic ingredients for further iterations. Namely, we
consider the reduction ${\cal L}^{(0)}\to {\cal L}^{(1)}$, where the index $%
n=0$ stands for an elementary set of interacting particles forming clusters.

$$
{} 
$$

Let $x_{j(\alpha )}$ be the position of the particle $j$ (integer) in the
cluster $\alpha $ (integer) where $1<\alpha <N^{(1)}$. Consider the
transformation to the center of mass $y_\alpha $, of the cluster $\alpha $,
given by 
\begin{equation}
x_{j(\alpha )}=y_\alpha +r_{j(\alpha )} 
\end{equation}
where $r_{j(\alpha )}$ denotes the relative distance with respect $y_\alpha $%
. Consider the interacting internal potential $U_T$, 
\begin{equation}
U_T=\sum q_{j(\alpha )}^{k(\beta )}U(x_{j(\alpha )}-x_{k(\beta )}), 
\end{equation}
where $q$ is a coupling parameter and the summation rules on all indices and
no self-interactions, or repeated indices, are assumed. Moreover, like $%
j(\alpha )$, the term $k(\beta )$ denotes the particle $k$ in the cluster at 
$y_\beta $. From (10), the potential transforms like 
\begin{equation}
U_T=\sum q_{j(\alpha )}^{k(\beta )}\left( U(y_\alpha -y_\beta )+(r_{j(\alpha
)}-r_{k(\beta )})U^{\prime }(y_\alpha -y_\beta )\right) +F(r) 
\end{equation}
where a first order multipolar expansion, in the internal coordinates, was
assumed and the symbol $U^{\prime }$ denotes the first derivative. Moreover
in (12), $F(r)$ denotes the linear terms depending only on the relative
coordinate and related to internal interaction in the cluster. Now we define
the coupling parameter, between clusters, $Q_\alpha ^\beta $ and the moment $%
d_{\alpha ,\beta }$ like 
\begin{equation}
Q_\alpha ^\beta =\sum_{j,k}q_{j(\alpha )}^{k(\beta )},\quad d_{\alpha ,\beta
}=\sum_{j,k}q_{j(\alpha )}^{k(\beta )}(r_{j(\alpha )}-r_{k(\beta )}). 
\end{equation}
and from equation (12) and (13) the internal interacting potential becomes 
\begin{equation}
U_T=\sum_{\alpha >\beta }Q_\alpha ^\beta U(y_\alpha -y_\beta )+d_{\alpha
,\beta }U^{\prime }(y_\alpha -y_\beta )+F(r). 
\end{equation}

$$
{} 
$$

Similarly, for an external field acting on every particle and given by 
\begin{equation}
V_T=\sum_{\alpha ,j}q_{j(\alpha )}V(x_{j(\alpha )}) 
\end{equation}
and defining the new coupling parameters and dipolar distribution by 
\begin{equation}
Q_\alpha =\sum_{j(\alpha )}q_{j(\alpha )},\quad m_\alpha =\sum_{j(\alpha
)}q_{j(\alpha )}r_{j(\alpha )}, 
\end{equation}
then the potential (15) can be written, at first order in the relative
coordinates, as

\begin{equation}
V_T=\sum_\alpha Q_\alpha V(y_\alpha )+m_\alpha V^{\prime }(y_\alpha )
\end{equation}
and like (12) the symbol $V^{\prime }$ denotes the first derivative. Remark
that no term like $F(r)$ appears in this case. From (14,17), and the
transformation for the kinetic term which is form invariant, the complete
Hamiltonian becomes

\begin{equation}
H=H_S+H_R+H_I 
\end{equation}
where the explicit form for the Hamiltonians are 
\begin{equation}
H_S=\sum_\alpha \frac{p_\alpha ^2}{2\mu _\alpha }+Q_\alpha V(y_\alpha
)+\sum_{\alpha >\beta }Q_\alpha ^\beta U(y_\alpha -y_\beta ) 
\end{equation}
\begin{equation}
H_I=\sum_\alpha ^{}V^{\prime }(y_\alpha )m_\alpha +\sum_{\alpha >\beta
}U^{\prime }(y_\alpha -y_\beta )d_{\alpha ,\beta } 
\end{equation}
and $H_R$ is the contribution due only to internal coordinates $(r,\dot r)$.
In equation (19) the term $\mu _\alpha $ denotes the total mass of the
cluster $\alpha $. Recall that it was always assumed a first multipolar
order expansion in the internal coordinate.

$$
{} 
$$

In this way, the Hamiltonian (18) has the structure worked in section I.
Since the internal variables $d_{\alpha ,\beta }$ and $m_\alpha $ are
assumed independent and they have the loss-memory property then , from
section I, the evolution operator for the ensemble of mesoparticles with
position $y_\alpha $ is

\begin{equation}
{\cal L}^{(1)}={\cal L}^{(0)}(y,p)+\sum_\alpha \gamma _\alpha ^{(1)}\left( 
{\cal L}_{IV}^{(1)}(y_\alpha )\right) ^2+\sum_{\alpha >\beta }\gamma
_{\alpha ,\beta }^{(1)}\left( {\cal L}_{IU}^{(1)}(y_\alpha -y_\beta )\right)
^2 
\end{equation}
where we have for every component the expressions : 
\begin{equation}
{\cal L}^{(0)}(y,p)\circ =\left\{ H_S,\circ \right\} 
\end{equation}

\begin{equation}
\left( {\cal L}_{IV}^{(1)}(y_\alpha )\right) ^2\circ =\left\{ V^{\prime
}(y_\alpha ),\left\{ V^{\prime }(y_\alpha ),\circ \right\} \right\} 
\end{equation}

\begin{equation}
\left( {\cal L}_{IU}^{(1)}(y_\alpha -y_\beta )\right) ^2\circ =\left\{
U^{\prime }(y_\alpha -y_\beta ),\left\{ U^{\prime }(y_\alpha -y_\beta
),\circ \right\} \right\} 
\end{equation}
namely a double Poisson brackets, or double commutator in the quantum case.

$$
{} 
$$

In (21) the parameters $\gamma $ are related to white-noise type
correlations between internal variables (dipole moments) for every
mesoparticle . This is the Markovian approximation where memory effects are
ignored. Expression (21) gives us the evolution operator ${\cal L}^{(1)}$
for $N^{(1)}$ mesoparticles of coordinates $(y_\alpha ,p_\alpha )$, where
internal degrees of freedom were eliminated. We noticed that assumption
(iii), of section I, is in accord with our deduction because the evolution
operator at scale $n=1$ was deducted from this one at scale $n=0$. Similar
equations for one, or two mesoparticles, were also discussed in [28-30]. The
idea to use internal degrees of freedom as an internal environment are also
discussed in [31].

$$
$$

The above evolution operator (21) is related to the first elimination of
internal degrees. Nevertheless,  if we can recognize a second structure of
clusters, we can eliminate new internal degrees. Evidently, this will be
possible only if the Markovian approximation is valid. Let ${\cal L}%
^{(n)}(x_j,q_j)$ be the evolution operator at scale $n$, which includes
kinetics and potential terms like (9), and consider a cluster recognition
with center of masses at $y_\alpha $ where $1<\alpha <N^{(n+1)}$. Then the
formal first order multipolar expansion with respect to the relative
coordinates (10) is

\begin{equation}
{\cal L}^{(n)}={\cal L}^{(n)}(y,Q)+\sum_\alpha \frac{\partial {\cal L}%
^{(n)}(y,Q)}{\partial y_\alpha }d_\alpha +F(r) 
\end{equation}
where the formal derivative stands for the first order expansion, and $%
d_\alpha $ are linear function of the internal degrees. Moreover, $Q$
denotes some re-defined coupling parameters. Then still we have a situation
similar to this of section (I), and if we assume the loss memory effect then
the new evolution operator for the $N^{(n+1)}$ mesoparticles will be

\begin{equation}
{\cal L}^{(n+1)}={\cal L}^{(n)}(y,Q)+\sum_\alpha \gamma _\alpha ^{(n)}\left( 
\frac{\partial {\cal L}^{(n)}(y,Q)}{\partial y_\alpha }\right) ^2. 
\end{equation}

$$
{} 
$$

At this point some remarks related to the iterating equation (26) are: (a)
The formal derivative stands for the first order multipolar expansion around
the center of mass of every mesoparticle, or cluster. (b) Expression (26) is
valid in the quantum or classical case, where ${\cal L}$ becomes related to
a set of elementary commutators or Poisson brackets. (c) Space structure
(lattice, fractal, disordered, etc.) is contained in ${\cal L}^0$. It
decides the criterion for the cluster recognition. (d) Remark that the free
particles case (${\cal L}_V={\cal L}_U=0$) is a trivial
form-invariant-example under reduction procedure.

$$
$$

\begin{section}*{III.- Examples}
\end{section}
$$
$$

We will now examine briefly some examples related to the reduction procedure
discussed in section II. Explicitly, we consider an ensemble of harmonic
interacting particles with anharmonic external fields. In fact, the
structure of the evolution operator becomes invariant after some reductions
(aside of some redefined coupling parameters). The search for such an
invariance was also investigated in [3] for a two-parameter model of
decoherence, where the equation of motion for the center of mass, is
formally identical to the equation for the microscopic constituents. 
$$
{} 
$$

Consider an ensemble of particles with harmonic interaction, in a nonlinear
external field. The internal and external interaction operators are given by 
\begin{equation}
{\cal L}_U^{(0)}\circ =\sum_{i\neq j}(1/2)\{K_{i,j}(x_i-x_j)^2,\circ \}. 
\end{equation}

\begin{equation}
{\cal L}_V^{(0)}\circ =\lambda \sum_j\left\{ x_j^3,\circ \right\} 
\end{equation}
where $K_{i,j}$ are positive constants and $\lambda $ a coupling parameter.
The expansion around $N^{(1)}$ center of mass gives 
$$
{\cal L}_U^{(0)}\circ =\sum_{\alpha \neq \beta }(1/2)\{K_{\alpha ,\beta
}^{\prime }(y_{j(\alpha )}-y_{k(\beta )})^2,\circ \}+ 
$$
\begin{equation}
+\sum_{\alpha \neq \beta }\sum_{j,k}\{K_{j(\alpha ),k(\beta )}(y_\alpha
-y_\beta )(r_{j(\alpha )}-r_{k(\beta )}),\circ \}+F(r) 
\end{equation}
and for the external component 
\begin{equation}
{{\cal L}_V^{(0)}\circ =\lambda ^{\prime }\sum_\alpha \left\{ y_\alpha
^3,\circ \right\} +3\lambda \sum_{\alpha ,j}\left\{ y_\alpha ^2r_{j(\alpha
)},\circ \right\} } 
\end{equation}
where $K^{\prime }$ and $\lambda ^{\prime }$ are re-defined coupling
parameters. The reduction procedure (section II) leads to the evolution
operator 
$$
{\cal L}^{(1)}\circ =\sum_\alpha \left\{ \frac{p_\alpha ^2}{2\mu _\alpha }%
+\lambda ^{\prime }y_\alpha ^3,\circ \right\} +\sum_{\alpha ,\beta
}K_{\alpha ,\beta }^{\prime }\left\{ \left( y_\alpha -y_\beta \right)
^2,\circ \right\} + 
$$
\begin{equation}
+\sum_{\alpha ,\beta }\gamma _{\alpha ,\beta }^K\left\{ \left( y_\alpha
-y_\beta \right) ,\left\{ \left( y_\alpha -y_\beta \right) ,\circ \right\}
\right\} +\sum_\alpha \gamma _\alpha ^\lambda \left\{ y_\alpha ^2,\left\{
y_\alpha ^2,\circ \right\} \right\} 
\end{equation}
which gives us the evolution equation for the set of mesoparticles.

$$
{} 
$$

At this point we have an interesting result, a second reduction process,
makes invariant the (internal) evolution operator. In fact, the only changes
are related to the redefinition of the coupling parameters and mass. The
same is true for the external anharmonic term in (31), which becomes
invariant after three reduction process. This seems a general fact related
with the formal derivative in the expression (26) for the evolution
operator. So, for systems interacting algebraically (i.e. $V,U\sim x^n$) the
reduction procedure seems invariant after a number finite of steps. Namely,
the laws of evolution become the same at different scales of perceptions.
Nevertheless, the reduction procedure must be stopped when there is not
loss-memory-effect and then this process does not can be continuously
carried-out. This is the case for instance for a system with a finite number
of constituent. Also, we noticed that geometric aspects must be considered
at every reduction, and some interesting candidates for a such invariance
are elementary excitations in fractal structures. 
$$
{} 
$$

\begin{section}*{IV.- Wigner function and the  classsical limit}

\end{section}

Since (21) is also valid for quantum system making the appropriate changes,
it is instructive to study the semiclassical limit using the Wigner
function. In fact, we shall find that the semiclassical limit needs some new
conditions because the decoherence term related to the parameter $\gamma $.

$$
{} 
$$

The Wigner function $\rho _w$ defined by the Fourier transform of the
density operator in coordinate representation is given by

\begin{equation}
\rho _w(x,p,t)=\frac 1h\int d\eta e^{ip\eta /\hbar }\rho (x-\eta /2,x+\eta
/2,t). 
\end{equation}
Where $\rho (x,y,t)$ is the statistical operator in coordinate
representation. Then, from equation (21) and keeping by simplicity only the
external potential term, the evolution for the Wigner function becomes

$$
\partial _t\rho _w=\left\{ H_S,\rho _w\right\} +\gamma \left\{ \frac{%
\partial V}{\partial x},\left\{ \frac{\partial V}{\partial x},\rho
_w\right\} \right\} + 
$$
\begin{equation}
+(\hbar ^2/24)\left( \frac{\partial ^3V}{\partial x^3}\frac{\partial ^3\rho
_w}{\partial p^3}-2\gamma \frac{\partial ^2V}{\partial x^2}\frac{\partial ^4V%
}{\partial x^4}\frac{\partial ^4\rho _w}{\partial p^4}\right) +O(\hbar ^4). 
\end{equation}
The first two terms are the classical operators (22-23) and the other are
related to the quantum contribution. So, the mesoscopic term related to the
parameter $\gamma $ gives new quantum corrections. The usual semiclassical
approximation, when $\gamma =0$, is the well-known relationship

\begin{equation}
\left| \frac{\partial V}{\partial x}\frac{\partial \rho _w}{\partial p}%
\right| \gg \hbar ^2\left| \frac{\partial ^3V}{\partial x^3}\frac{\partial
^3\rho _w}{\partial p^3}\right| 
\end{equation}
and related roughly to the optical geometrical limit. It must be noted,
however, a similar condition related to the mesoscopic term from (33) this
condition is ($\gamma \neq 0$) 
\begin{equation}
\left| \left( \frac{\partial ^2V}{\partial x^2}\right) ^2\frac{\partial
^2\rho _w}{\partial p^2}\right| \gg \hbar ^2\left| \frac{\partial ^2V}{%
\partial x^2}\frac{\partial ^4V}{\partial x^4}\frac{\partial ^4\rho _w}{%
\partial p^4}\right| . 
\end{equation}

$$
{} 
$$
Namely, it explores even derivatives in the potential.

$$
{} 
$$

Finally we note that the deduction of the evolution equation (33), for the
Wigner function, was obtained assuming (integration by part) 
\begin{equation}
i\hbar \partial _x\rho (x-\eta ,x+\eta )e^{ip\eta /\hbar }|_{\eta =0}^{\eta
=\infty }=0. 
\end{equation}
Namely, the vanishing of the correlation term at infinite. This requirement
is not always verified, for instance, consider the states $\rho =\psi
(y)^{*}\psi (x)$ where the wave function is $\psi =\sin kx$ which does not
satisfied (36). Nevertheless, the contribution due to the decoherence in
(21), produces a fast annulment of the off-diagonal terms in the statistical
operator for short-range external potential. In this way, the condition (36)
can be satisfied for times greater that the decoherence time, and given a
solid support to the evolution equation (33) for the Wigner function.

$$
{} 
$$

\begin{section}*{ V.- Conclusions and discussions} 
\end{section}

$$
$$

We have considered a hierarchy of master equation describing the evolution,
at different scales of perception, for ensembles of mesoparticles.
Explicitly, the equation (26) gives us formally the evolution operator $%
{\cal L}^{(n+1)}$ from this one at scale $n$ (assumption (iii)). Its
deduction requires a systematic coordinate change to the centers of mass,
defined by some physical constraints, and the elimination of internal
degrees of freedom is carried-out assuming loss-memory-effects. This
Markovian approximation is not always valid and then, in such a case, the
reduction process must be stopped. The first reduction procedure was
carried-out explicitly for an ensemble of elementary components (21) 
$$
{} 
$$

It must be noted that assumptions (i-iii), of section I, are the basis where
our reduction procedure was developed. Namely, they possibility the
obtention of a hierarchy of master equations at different scales of
perception for cluster of mesoparticles. Some similarities between (i-iii)
and those used in the architecture of complex synthetic assembles would be
found in reference [19,20].

$$
{} 
$$

On the other hand, some important points related to the deduction of the
evolution equation (21) and (26) are :

$$
{} 
$$

(a) The first multipolar order expansion, in the interaction term, tell us
that the asymptotic limit $t\to +\infty $ must be carried-out carefully at
different scales [29].

(b) Decoherence effects at every discrete scale appears usually related to
the reduction technique (section I). Thus, decoherence at macroscopic level
is due to the internal complexity of every macroparticle. And quantum
superposition is turned into statistical mixture .

(c) The search for invariant systems was carried-out explicitly for a model
composed of interacting oscillators with an anharmonic term. It seems that
other invariant systems could be found.

(d) It was assumed that the internal moments, like to $d$ or $m$ in (13) and
(16), are random and independent. Obviously this is not easy to prove and we
have only assume that behavior. The statistical independence, between these
random variables, is a simplification related with our calculations.

(e) The deduction of a general equation like (7) was carried-out assuming
special initial conditions ($\rho (t=0)=\rho _R^e\rho _S$). In our specific
case of section II, these conditions not necessarily hold. More explicitly,
the internal interaction cannot be switched-up arbitrary. It is an open
problem to prove the validity of our procedure in this case.

$$
{} 
$$

To explore possible applications of our prescription, we can consider trends
like wavematter, currently studied theoretically as well experimentally.
After all, interacting atoms in external fields could be considered as
mesoparticles. Specifically, with laser cooling techniques it becomes
possible to cool atoms so that the quantum nature of atomic center of mass
motion becomes important [32,33]. Also, it can be interesting note the
growing interest in new mechanisms to break Anderson localization in
disordered systems [34]. Particularly, there is the controversy about the
possible enhancement of the localization length for interacting particles
(TIP) [35]. This suggests considering the behavior of mesoparticles in
disordered systems. Namely, an equation like to (21) with external random
potential. After all, localization is a phenomenon related to coherence
which is loss due to internal complexity for mesoparticles. A more detailed
treatment of these points, and further physical applications will be given
elsewhere.

$$
$$

\begin{section}*{Acknowledgments} 
\end{section}

This work was partially supported by grant FONDECYT 394 000 4. It was made
possible due to the influence which I received from the group of dynamical
system directed by professor E. Tirapegui (FCFM).

$$
{} 
$$

\end{document}